\documentclass[prb,twocolumn,showpacs,preprintnumbers,amsmath,amssymb]{revtex4}
\usepackage{amsmath,amssymb,amsfonts}
\usepackage{bm}
\usepackage{graphicx}
\usepackage{times}
\usepackage{color}
\usepackage[colorlinks,bookmarks=false,citecolor=blue,linkcolor=red,urlcolor=blue]{hyperref}

\newcommand{\fref}[1]{Fig.~\ref{#1}}
\newcommand{\eref}[1]{Eq.~(\ref{#1})}

\newcommand{\normwidth}{0.8\columnwidth}

\begin{document}

\title{Surface Polaron Formation in the Holstein model}
\author{Reza Nourafkan}
\affiliation{Department of Physics, Sharif University of Technology,
P.O.Box: 11155-9161, Tehran, Iran}
\author{Massimo Capone}
\affiliation{SMC, CNR-INFM and Dipartimento di Fisica, Universit\`a Sapienza, P.le Aldo Moro 2, I-00185, Roma, Italy}
\author{Nasser Nafari}
\affiliation{Institute for Research in Fundamental Sciences (IPM),
19395-5531, Tehran, Iran}

\begin{abstract} The effect of a solid-vacuum interface on the properties of a strongly coupled electron-phonon
system is analyzed using dynamical mean-field theory to solve the Holstein model in a semi-infinite cubic lattice.
Polaron formation is found to occur more easily (i.e., for a weaker electron-phonon coupling)
 on the surface than in the bulk. On the other hand, the metal-insulator transition associated to
 the binding of polarons takes place at a unique critical strength in the bulk and at the surface.
\end{abstract}
 \maketitle

\section{Introduction}
Convincing experimental evidence of polaronic behavior has been reported in materials such as the high-$T_c$ cuprates
and manganites.  For instance, the transition from the low temperature ferromagnetic metallic state to the paramagnetic insulating state in manganites is caused by the formation of combined structural/magnetic polarons\cite{Lynn}. Signatures of small polarons  have been observed in undoped cuprates \cite{optics,Shen}.

From a theoretical point of view, polaron formation has been
intensively studied using a number of approaches. An important role
to improve our understanding of polaronic phenomena has been
recently played by Dynamical Mean-Field Theory (DMFT)\cite{Georges}.
DMFT is a powerful non-perturbative tool for strongly interacting
systems. This technique, which becomes exact in the limit of
infinite coordination number, reduces the full lattice many-body
problem to a local impurity embedded in a self-consistent effective
bath of free electrons.

DMFT studies of the half-filled Holstein model in a Bethe lattice with a semi-elliptic free
density of states have clarified the difference between the polaron crossover,
i.e., the continuous and progressive entanglement between electrons and phonons,
and the bipolaronic metal-insulator transition\cite{CaponeCiuchi,Capone3,Holstein_altri}.
If no symmetry breaking is allowed, for small e-ph couplings the ground state is metallic with Fermi liquid characteristic.
 Upon increasing the e-ph coupling, the carriers lose mobility, eventually acquiring polaronic character, with a finite
lattice distortion associated to the electron motion. Polaron formation occurs as a continuous crossover.
Once formed, polarons tend to attract and form a bound pair in real space, called bipolaron
\cite{CaponeCiuchi}. Within DMFT the bipolaronic binding gives rise to an insulating state of localized
pairs\cite{pairing_insulator}, and bipolaron formation gives rise to a metal-insulator transition.
The pairing transition does not coincide with the polaron crossover:
Polarons are formed before (i.e., for a weaker coupling) the pairing transition occurs as
long as the typical phonon frequency is smaller than the electronic
energy scales (adiabatic regime).

On the other hand, fabrication of a variety of hetrostructures and interfaces
involving cuprates and manganites raises the question of
whether the electronic behavior at the surface or interface is
different from the bulk. Several studies have been devoted to the case
of repulsive electron-electron interactions (Hubbard model) and to
a vacuum-solid interface.
Potthoff and Nolting \cite{Potthoff1}, and
Liebsch \cite{Liebsch} have argued that reduced coordination at the
surface may enhance correlation effects. They also studied the
magnetic ordering induced by enhanced correlation at the surface.
Matzdorf \textit{et al}.\cite{Matzdorf} proposed that ferromagnetic
ordering is stabilized at the surface by a lattice distortion.
Surface ferromagnetism had been also discussed in a dynamical mean
field theory of Hubbard model by Potthoff and Nolting
\cite{Potthoff2}. Helmes \textit{et al}. studied the scaling
behavior of the metallic penetration depth into the Mott insulator
near the critical Coulomb interaction within the Hubbard model
\cite{Helmes}. Borghi \textit{et al}. have shown the existence of a dead
surface layer with exponentially suppressed quasiparticles\cite{borghi}.

Here we concentrate on the effect of a vacuum-solid surface on interacting
electron-phonon systems, and we study the formation of polarons and the
transition to a bipolaronic insulating state in the semi-infinite
Holstein model at half-filling and zero temperature on the bipartite
simple cubic (sc) lattice with nearest-neighbor hopping. While the
occurrence of charge transfer is typical for a system with reduced
translational symmetry, in our model at half-filling any charge
transfer is excluded by the particle-hole symmetry, leading to a
homogeneous charge distribution among the layers parallel to the
surface, and local occupations near the surface do not differ from
the average filling, $\langle n_{\alpha} \rangle = \langle n \rangle
= 1$, where $\alpha$ labels each layer.

In addition to the geometrical effect of missing neighbors, the
surface electronic structure of interacting electron systems is also
complicated by the fact that the microscopic interactions in the
vicinity of the surface have values which may significantly differ
from those in the bulk. A relaxation of the surface layer, for
example, changes the overlap between the one-particle basis states
and thus implies a modified hopping integral. Within the Holstein
model, the parameter modifications will be reflected in
different values of the surface topmost layer hopping integrals and
e-ph coupling strengths relative to the bulk ones.
In this work we will not consider this effect, in order to focus on the
more intrinsic effects that can give rise to different physics in the surface.

This paper is organized as follows: In Sec. II the model is
introduced and the application of DMFT for surface geometry is
briefly discussed. We mainly characterized the electronic and
phononic properties by considering the layer quasiparticle weights,
double occupancies and the phonon probability distribution function.
The corresponding results are discussed in Sec. III. Finally
Sec. IV is devoted to concluding remarks.

\section{The Model and Method}

The Holstein Hamiltonian is defined by:
\begin{eqnarray}
\label{eq:Holstein}
H=&-&t\sum_{\langle ij \rangle \sigma}{\left( c^\dagger_{i\sigma}c_{j\sigma} + c.c. \right)}\nonumber \\
  &+& g\sum_{i}{\left( n_{i} - 1\right)\left( b^\dagger_{i} + b_{i} \right)} +
  \Omega_{0}\sum_{i} b^\dagger_{i}b_{i},
\end{eqnarray}
where $c_{i\sigma}\left(c^\dagger_{i\sigma} \right)$  and
$b_{i}\left(b^\dagger_{i} \right)$ are, respectively, destruction
(creation) operators for itinerant electrons with spin $\sigma$ and
local vibrons of frequency $\Omega_{0}=0.2t$ on site $i$, $n_{i}$ is
the electron density on site $i$, $t$ stands for the itinerant
electrons hopping matrix elements between the nearest-neighbor
sites, and $g$ denotes the  electron-phonon coupling. We fix the
energy scale by setting $t=1$.

To obtain the ground state properties of this model, we use the embedding
approach introduced by Ishida and Liebsch\cite{Ishida} to extend DMFT
to inhomogeneous systems. In this scheme, the system is divided into two parts:
The surface region which includes the first $N$ layers, and the
adjacent semi-infinite bulk region (substrate) which is coupled to
it. Next, we represent the effects of the substrate on the surface
region by a complex, energy-dependent, embedding potential acting on
the Hamiltonian matrix of the surface region.The embedding method requires to
consider a relatively small number of surface layers and it is therefore a
 computationally less expensive extension of DMFT in the presence of an
interface as compared to the slab method, in which the inhomogeneous system is
simply represented as a finite number of layers\cite{Ishida}.

Because of translational symmetry in the plane perpendicular to the interface,
the embedding potential of the substrate is
diagonal with respect to the two-dimensional wave vector ${\bf
k}=(k_x,k_y)$ and can be expressed as an $N\times N$ matrix by
\begin{equation}
{\bf S}({\bf k}, i\omega_n) = \tilde{{\bf T}} {\bf G}({\bf k}, i\omega_n) {\bf T},
\label{eq:embedding}
\end{equation}
where ${\bf G}({\bf k}, i\omega_n)$ is the Green's function of the
substrate defined by
\begin{equation}
{\bf G}({\bf k}, i\omega_n) = \big[(i\omega_n+\mu){\bm 1}-{\bm
\epsilon}({\bf k})-{\bf \Sigma}(i\omega_n) \big]^{-1}.
\label{eq:green}
\end{equation}
In here, ${\bf \Sigma}(i\omega_n)$ is the bulk self-energy, which in
the framework of single-site DMFT, is independent of wave vectors,
$\bf k$, and $\omega_n$ are the Matsubara frequencies. We obtain
the self-energy by performing a standard DMFT calculation for the
bulk crystal corresponding to the substrate. $\mu$ is the chemical
potential and ${\bm \epsilon}({\bf k})$ is the two-dimensional
dispersion relation, which includes information about surface
geometry. The ${\bm \epsilon}({\bf k})$ matrix for the surface cutting a simple cubic lattice
along the z direction [sc(001) surface] takes the following form \cite{Potthoff1}:
\begin{eqnarray}
{\bm \epsilon}({\bf k})=\left(
\begin{array}{cccc}
t\epsilon_{\parallel}({\bf k}) & t\epsilon_{\perp}({\bf k}) & 0 &0 \\
t\epsilon_{\perp}({\bf k}) & t\epsilon_{\parallel}({\bf k}) & t\epsilon_{\perp}({\bf k}) & 0 \\
 0 & t\epsilon_{\perp}({\bf k}) & t\epsilon_{\parallel}({\bf k}) & \cdots \\
0 & 0 & \cdots & \cdots
\end{array}
\right). \label{eq:epsilon}
\end{eqnarray}
The intralayer (parallel) hopping and the interlayer (perpendicular)
hopping are specified by $t\epsilon_{\parallel}({\bf k})$ and
$t\epsilon_{\perp}({\bf k})$, respectively, and are given by
\begin{equation}
\epsilon_{\parallel}=-2[\cos (k_x)+\cos (k_y)], \,\,\,\, \vert \epsilon_{\perp}({\bf {k}})\vert^{2}=1 .
\end{equation}
Finally, $\bf T$ is the hopping matrix between primitive cells of
substrate and surface region. Since $\bf T$ is non-zero between
nearest-neighbor layers of substrate and surface region, only the
surface Green's function \cite{Kalkstein} of the substrate need to
be considered in \eref{eq:embedding}.

After constructing the embedding potential of the substrate, ${\bf
S}({\bf k}, i\omega_n)$, by way of a coupled-layer DMFT calculation
in the surface region the self-energy matrix is determined
self-consistently. This can be achieved via the following steps: (i)
associating an effective impurity model with each layer in the
surface region, solving them by using an impurity solver to find the
layer-dependent local self-energies, $\Sigma_{\alpha}(i\omega_n)$,
and constructing the surface region self-energy matrix which is
diagonal in layer indices $(\alpha, \beta)$ with the elements,
$\Sigma_{\alpha
\beta}(i\omega_n)=\Sigma_{\alpha}(i\omega_n)\delta_{\alpha \beta}$,
(ii) calculating the on-site layer-dependent Green's function via the
following relation:
\begin{multline}
G_{\alpha}(i\omega_n)=\\ \sum_{{\bf k}}\left(
\frac{1}{(i\omega_n+\mu) {\bm 1}-{\bm \epsilon}({\bf k})-{\bf
S}({\bf k}, i\omega_n)-{\bm \Sigma}(i\omega_n)} \right)_{\alpha
\alpha}, \label{eq:LayerG}
\end{multline}
where $N\times N$ ${\bm \epsilon}({\bf k})$ matrix is given by
\eref{eq:epsilon}, (iii) implementing the DMFT self-consistency
relation for each layer, ${\mathcal G}^0_{\alpha}( i\omega_n) =
\big[G_{\alpha}^{-1}(i\omega_n) +
\Sigma_{\alpha}(i\omega_n)\big]^{-1}$, which determines the bath
parameters for the new  effective impurity model. The cycles have to
be repeated until self-consistency is achieved.

We use the exact diagonalization (ED) technique to solve the
effective impurity model at zero temperature\cite{Caffarel}, which
works equally well for any values of the parameters and only
involves a discretization of the bath hybridization function,
 which is described in terms of a finite and small set
of levels $n_s$ in order to limit the Hilbert space to a workable
size. For the case of phonon degrees of freedom we considered here,
the infinite phonon space is also truncated allowing for a maximum
number of excited phonons $N_{ph}$. The typical values we considered
for the bath levels are $n_s=8-9$ and typical maximum number of
phonons are $N_{ph}=30-50$. We tested that these numbers, indeed,
provide converged results. Moreover, the number of surface layers is
chosen to be $N=5$ in all calculations.

\section{Results}

\begin{figure}
  \begin{center}
    \includegraphics[width=\normwidth]{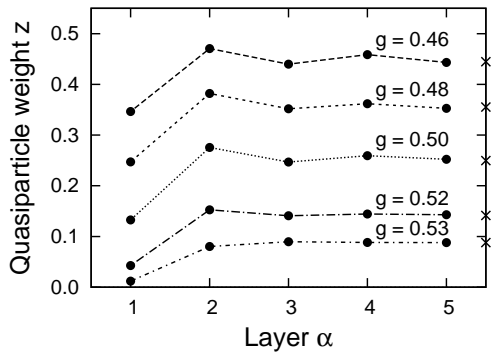}
    \caption{Quasiparticle weight $z$ of semi-infinite Holstein model for
    simple cubic lattice in the (001) orientation as a function of layer index $\alpha$.
    Crosses on the vertical axis on the right hand side indicate the bulk $z$ corresponding to five given-values of $g$.
     Lines are drawn as a guide to the eye. } \label{fig:fig1}
  \end{center}
\end{figure}

As we anticipated in the introduction, all our calculations are
performed for the case of uniform parameters. This assumption,
together with the half-filling condition which enforces charge
homogeneity, let us to single out the effect of the interface and to
focus on the purely geometrical aspect of the problem.
\fref{fig:fig1} shows the calculated quasiparticle weight
$z_{\alpha}$ of the semi-infinite Holstein model at $T=0$ in the
metallic range as a function of layer index $\alpha$, where the
outermost layer corresponds to $\alpha = 1$.
$z_{\alpha}$ measures the metallic nature of a system, $z$ being one for
a non-interacting metal and zero for a correlated insulator. In our
case $z=0$ implies a bipolaronic insulator.
The crosses on the vertical axis on the right hand side indicate the $z$ values of the
bulk metal determined by a separate bulk DMFT calculation.
For any value of the coupling, the quasiparticle weight of the
surface layer $z_{\alpha=1}$ is significantly reduced compared to
$z_{\alpha = 2}$ and $z_{\alpha = 3}$ which can be understood as the effect of the
reduced surface coordination number and enhanced effective
correlations, in complete analogy to the results for repulsive
interactions.
The evolution as a function of the layer
index depends instead on the coupling regime.
For weak and moderate coupling, $z_{\alpha}$ has a non monotonic behavior
  which is damped with increasing distance to the
surface. For $g$-values closer to the critical coupling strength of
the bulk bipolaronic transition, $g_{c}$ ($g_{c}\approx 0.55$) the behavior changes
qualitatively. Here the layer dependence becomes monotonic, and the
quasiparticle weight quickly approaches its bulk value with
increasing $\alpha$.

\begin{figure}
  \begin{center}
    \center{\includegraphics[width=\normwidth]{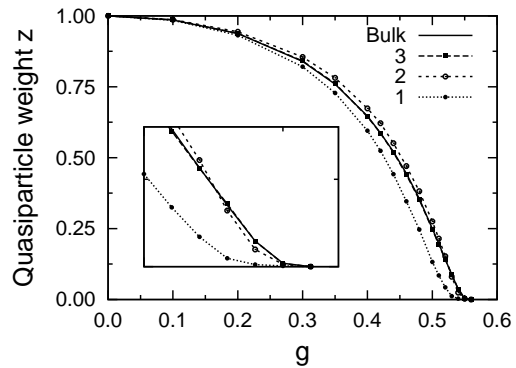}}\\
    \center{\includegraphics[width=\normwidth]{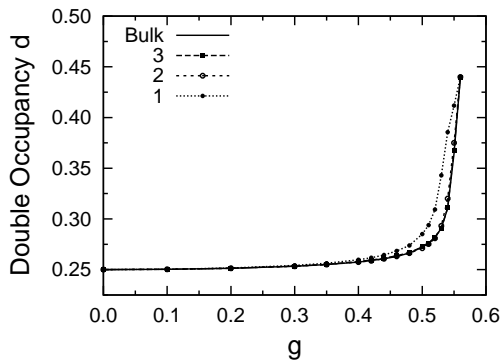}}
    \caption{Top panel : Layer-dependent quasiparticle weight $z_{\alpha}$ for the first three layers
of the semi-infinite Holstein model with simple cubic (001) surface
geometry and the bulk quasiparticle weight as a function of e-ph
coupling strength, $g$. $\alpha=1$ stand for the topmost surface
layer. The solid line shows $z$ for bulk calculations.
    The inset shows $z_{\alpha}(g)$ in the critical regime.
    Bottom panel : Layer-dependent double occupancy $d_{\alpha}$ as a function of $g$. }
 \label{fig:fig2}
  \end{center}
\end{figure}

\fref{fig:fig2} illustrates the layer-dependent
quasiparticle weight $z_{\alpha}$ for the first three layers and the bulk
quasiparticle weight as a function of $g$.
As expected, all the $z$'s monotonically decrease as a function of the e-ph
coupling, and they eventually vanish.
As can be seen in the figure,
the differences between the $z_{\alpha}$ and the bulk $z$ diminish with increasing
distance from the surface and for the third layer, the quasiparticle weight is almost
indistinguishable from the bulk $z$ on the scale used. It is crucial to observe that
the different $z_{\alpha}$ all vanish at {\it the same} value of $g$, which also
coincides with the bulk critical coupling strength, $g_{c}=g_{c,bulk}$.

If the surface and the bulk were decoupled, the reduced surface coordination
number would tend to drive the surface
to an insulating phase at a coupling strength lower than the bulk
critical coupling $g_c$. However, below $g_c$
the bulk excitations, due to hopping processes between the surface
and the bulk can induce a quasiparticle peak with a
non-zero weight $z_{\alpha=1}>0$ in the topmost layer and a real
surface transition is not found; $z_{\alpha=1}$ remain non-zero,
although being very small, up to the critical coupling for bulk
transition, $g_{c}$. The investigation of the imaginary part of the
layer-dependent self-energies, $\Sigma_{\alpha}(i\omega_n)$ at
$\omega_n \rightarrow 0$ (not shown) confirms the uniqueness of the
critical strength $g_{c}$. In the limit of $\omega_n \rightarrow 0$
and for all $g<g_c$, the imaginary part of self energy vanishes for
all layers as it must be for a Fermi liquid. In the coupling
constants close to $g_c$ and in the metallic regime, a significant
layer dependence of $Im \Sigma_{\alpha}(i\omega_n)$ for $\omega_n
\rightarrow 0$ with a considerably larger slope in the first layer
($\alpha=1$) is seen, which reflects the enhanced correlation effects
at the surface.
In the insulating state, $Im\Sigma_{\alpha}(i\omega_n)$ diverges for $\omega_n \rightarrow 0$.
Therefore, there is a unique critical strength $g_{c}$ at which all
quasiparticle weight functions, $z_{\alpha}(g)$, simultaneously
approach zero.

The layer-dependent average double occupancy, $d_{\alpha}=\langle
n_{\alpha \uparrow} n_{\alpha \downarrow}\rangle$, is shown in the
bottom panel of \fref{fig:fig2} as a function of $g$. For small $g$
and all layers, $d_{\alpha}$ increases gradually. At $g=g_{c}$ it
rapidly reaches $\approx 1/2$. In the metallic region, the double
occupancies are increased more rapidly at the topmost surface layer
as compared with the interior of the system. Again, this is due to
the stronger effective e-ph interaction which results from the
narrowing of the non-interactiong density of states at the surface.

We have thus far established that even in the presence of a surface,
the half-filled Holstein model undergoes a single bipolaronic
metal-insulator transition, despite the surface is less metallic than
the bulk for any $g < g_c$.
We now discuss how the surface influences the local lattice distortions,
measured by the phonon probability distribution function (PDF), $P(x)= \langle
\phi_0 \vert x \rangle \langle x \vert \phi_0\rangle$, where $\vert
x \rangle \langle x \vert$ is the projection operator on the
subspace for which the phonon displacement $\hat{x}$ has a given
value $x$, and $|\phi_0 \rangle$ is the ground state vector.
This quantity can be used to characterize the polaron crossover\cite{CaponeCiuchi_manga}.


In the absence of e-ph interaction $P(x)$ is a Gaussian centered around $x=0$.
A small e-ph coupling slightly broadens the distribution which remains centered around $x=0$,
implying that the coupling is not sufficient to give rise to a finite polarization of the lattice.
Continuously increasing the interaction one eventually obtains a bimodal distribution with two
identical maxima at $x=\pm x_0$. Those maxima are indeed associated with empty and doubly
occupied sites, and testify the entanglement between the electronic state and the lattice distortion,
which is precisely the essence of the polaron crossover. Thus, the appearance of a bimodal shape
in $P(x)$ is a marker of the polaron crossover\cite{CaponeCiuchi_manga,Capone3}.

\begin{figure}
  \begin{center}
    \center{\includegraphics[width=\normwidth]{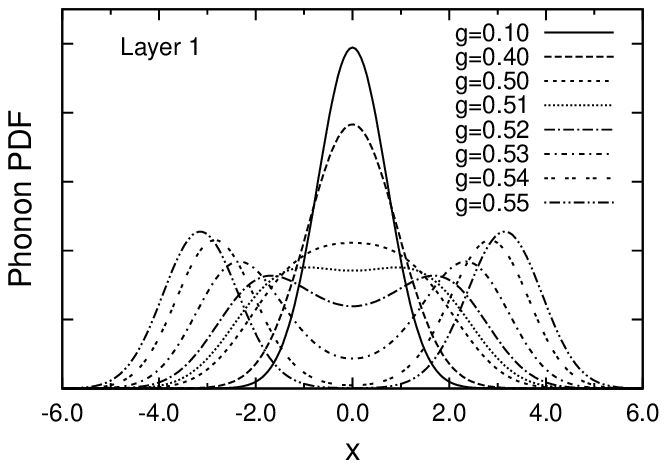}}\\
    \center{\includegraphics[width=\normwidth]{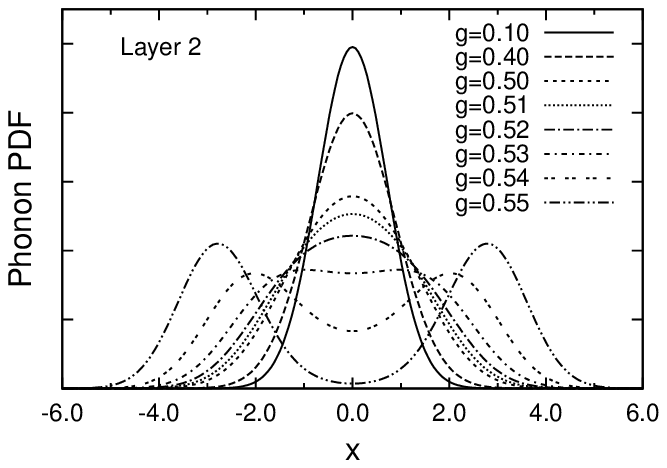}}
    \caption{Phonon probability distribution function for the first and second
    surface layers of a semi-infinite sc(001) Holstein model at half-filling.
    The various curves refer to different values of electron-phonon coupling strength, $g$.
    Upon increasing the e-ph coupling, a smooth crossover occurs between a unimodal
    distribution and a bimodal distribution. However, the polaron crossover (onset of bimodality)
    occurs at different values of $g$ for the first and second layers. Top panel shows that at the
    topmost surface layer, polaron formation takes place at $g\approx 0.51$, whereas at the second
    layer, it takes place at $g\approx 0.53$ (bottom panel). All other layers behave just like the second layer.
}
 \label{fig:fig3}
  \end{center}
\end{figure}

 \fref{fig:fig3} shows the polaron crossover for our semi-infinite Holstein
 model. For each layer the evolution as a function of the coupling follows
 the pattern we described above: The anharmonicity due to e-ph
interaction increases with increasing coupling strength leading
first to a non-Gaussian and finally to a bimodal PDF at all $g >
g_{pol}$ . This behavior signals the appearance of static
distortions, even if we are neglecting any ordering between them.
 The strongest differences with respect to the bulk PDF are
found for the top layer ($\alpha=1$) PDF. The layer PDFs converge to
the bulk PDF with increasing distance to the surface. Beyond the third layer the PDF
is essentially identical to its bulk behavior.
It is apparent from the data of  \fref{fig:fig3} that the PDF at the topmost (surface) layer
becomes bimodal at lower values of the
coupling strength with respect to the internal layers.
the surface can display polaronic distortions while the bulk is still
undistorted (even if the local vibrations are strongly anharmonic).

\section{Concluding Remarks}
We have investigated polaron formation and
transition to the bipolaronic insulating state at solid-vacuum surface
at zero temperature in the
framework of the semi-infinite Holstein model at half-filling.
Using the embedding approach to extend dynamical mean-field theory to
layered systems, it is found that the bipolaronic insulating state
occurs simultaneously at the surface and in the bulk, and it
takes place exactly at the same critical coupling strength $g_{c,bulk}$
as for the infinitely extended system, $g_{c,bulk}=g_{c}$.
When the system is metallic the
topmost layer quasiparticle weight $z_1$ is smaller than the bulk
value $z_{bulk}$, since a reduced surface coordination number
implies a stronger effective correlation effects.
Fixing the coupling at values quite  smaller than $g_c$, the quasiparticle weight is an oscillating
function of the layer index. As the distance from the
surface increases, these oscillations fade away.
For couplings close to the metal-insulator transition $z_{\alpha}$ instead
monotonically increases by approaching the bulk.
On the other hand, the polaron crossover occurs more easily at the surface with
respect to the bulk. There is therefore a finite window of e-ph coupling in which
the surface presents polaronic distortions, while the bulk has no distortions.
As we already mentioned, this difference is not able to support a metallic bulk
coexisting with an insulating surface.

Surface effects are expected to be the more pronounced the larger is
the number of missing neighbors in the topmost layer. As we move from the
sc(001) to the sc(011) and to the sc(111) surface geometries, the
surface coordination numbers decrease from $n^{(001)}_c=5$ to
$n^{(011)}_c=4$ to $n^{(111)}_c=3$, respectively. Therefore, we
expect to observe a narrowed topmost layer free density of state
which results in an enhanced ratio between the e-ph coupling
strength and the effective band width. Consequently, the e-ph
interaction tends to be stronger at the surface. Clearly, according
to this argument we expect the difference between the two coupling strengths
for the polaron formation at the surface and in the bulk would
become larger.

The present study has been restricted to uniform model parameters.
This leaves several open questions like the possibility of coexisting
different surface and bulk phases, if the model parameters at the
vicinity of the surface are modified (See the comment made in this
respect in Sec. I).


\begin{acknowledgements}
M.C. acknowledges financial support of MIUR PRIN 2007 Prot. 2007FW3MJX003
\end{acknowledgements}

\bibliographystyle{prsty}

\end{document}